\documentclass{pasa}%

\usepackage{mathrsfs}
\usepackage{breqn}
\newcommand{\HIsub}{\textnormal{\small \textsc{hi}}}

\title[HI Beyond the Local Universe]{Tracing HI Beyond the Local Universe}
\author[Meyer et al.]{Martin Meyer$^{1}$\thanks{martin.meyer@uwa.edu.au}, Aaron Robotham$^{1}$, Danail Obreschkow$^{1}$, Tobias Westmeier$^{1}$, Alan Duffy$^{2}$, \\Lister Staveley-Smith$^{1}$\\
\affil{$^1$International Centre for Radio Astronomy Research (ICRAR), The University of Western Australia, 35 Stirling Highway, Crawley, WA 6009, Australia}%
\affil{$^2$Centre for Astrophysics and Supercomputing, Swinburne University of Technology, PO Box 218, Hawthorn, VIC 3122, Australia}}%
\jid{PASA}
\doi{10.1017/pas.\the\year.xxx}
\jyear{\the\year}

\usepackage[authoryear]{natbib}
\bibpunct{(}{)}{;}{a}{}{,}
\usepackage{aas_macros}
\usepackage{hyperref} 
\hypersetup{colorlinks,citecolor=blue,linkcolor=blue,urlcolor=blue}

\begin{document}

\begin{abstract}
The SKA and its pathfinders will enable studies of HI emission at higher redshifts than ever before.  In moving beyond the local Universe, this will require the use of cosmologically appropriate formulae that have traditionally been simplified to their low-redshift approximations.  In this paper, we summarise some of the most important relations for tracing HI emission in the SKA era, and present an online calculator to assist in the planning and analysis of observations (\url{http://hifi.icrar.org}).

\end{abstract}

\begin{keywords}
radio lines: galaxies --  galaxies: distances and redshifts --- cosmology
\end{keywords}

\maketitle

\section{Introduction}

Neutral atomic hydrogen (HI) is one of the most important tracers for studying the assembly of mass, angular momentum, and structure in the Universe \citep[][]{lss2015,blyth2015,kim2015,kim2016,obreschkow2016}.  However, the difficulty of observing emission from this material has meant that most studies have been restricted to the local Universe, with only a small fraction of detections occurring beyond redshift z$\sim$0.1 \citep{jaffe2013,giovanelli2015}.  As powerful new radio telescopes come on-line (ASKAP, MeerKAT, WRST-Apertif, FAST, SKA), such observations will become commonplace \citep[][]{duffy2012,holwerda2012,lss2015}, necessitating the use of formulae that allow for an evolving Universe.  A basic summary of the cosmologically appropriate formulae for a variety of standard HI quantities is given in the following sections, with the aim of providing a reference document for upcoming pathfinder studies.  In the final section, we introduce an accompanying online calculator based on the provided formulae. Useful references for work underpinning the material here include \citet{hogg1999}, \citet{abdalla2005}, \citet{obreschkow2009}, Ned Wright's online tutorial\footnote{\url{www.astro.ucla.edu/~wright/cosmo_01.htm}}, and CosmoCalc.\footnote{\url{cosmocalc.icrar.org}}

\section{Redshift}

For observational HI studies, the redshift of a source is simply defined as:
\begin{eqnarray}
z_{\rm obs} = \frac{\nu_{\HIsub}}{\nu_{\rm obs}} - 1 = \frac{\lambda_{\rm obs}}{\lambda_{\HIsub}} - 1\,,
\end{eqnarray}

\noindent where $\nu_{\HIsub}$ and $\lambda_{\HIsub}$ are the emitted frequency and wavelength of the HI line, and $\nu_{\rm obs}$ and $\lambda_{\rm obs}$ are the corresponding frequency and wavelength at which it is observed.  This redshift is the combined result of a number of potentially contributing effects, including the cosmological redshift due to the expansion of the Universe, local motions in either the source or observer rest frames, and gravitational redshifts caused by (evolving) potential wells along the line of sight.  Considering just the cosmological redshift ($z_{\rm cos}$) and the peculiar motions of the source ($z_{\rm pec}^{\rm source}$) and observer ($z_{\rm pec}^{\rm observer}$) \citep{davis2014}:
\begin{eqnarray}
1+z_{\rm obs} = (1+z_{\rm cos})(1+z_{\rm pec}^{\rm source})(1+z_{\rm pec}^{\rm observer}) \,.
\label{eqn:zobs_wzpecobserver}
\end{eqnarray}

Different standards of rest or flow models are often used to minimise peculiar velocity effects, particularly in the nearby Universe.  At higher redshifts \citep[e.g. $z\gtrsim0.03$,][]{baldry2012}, the cosmic microwave background offers a reference frame that can be used to remove local peculiar velocity effects, reducing Equation~\ref{eqn:zobs_wzpecobserver} to:
\begin{eqnarray}
1+z_{\rm obs} = (1+z_{\rm cos})(1+z_{\rm pec}^{\rm source}) \,.
\label{eqn:zobs}
\end{eqnarray}

In this paper, unless otherwise specified, we make the overall simplifying assumption that the observed redshift is  equal to the cosmological redshift ($z_{\rm obs}$ = $z_{\rm cos} = z$, i.e. $z_{\rm pec}^{\rm observer}$ = $z_{\rm pec}^{\rm source}=0$), but be aware that, particularly in the local Universe, there can be a non-negligible difference between these redshifts and you may need to adjust for this assumption.  

\section{Parametrizations of Cosmic Expansion}

The relationship between proper (physical) distance and the recessional velocity of galaxies due to the expansion of the Universe at a given epoch is parametrized through the Hubble parameter $H(z)$, with Hubble's Constant $H_{\rm 0} = H(0)$ its value at the present time.  If we divide the energy density of the Universe, normalised by the critical energy density, into its fractional components of matter ($\Omega_{\rm M}$), radiation ($\Omega_{\rm R}$), vacuum energy ($\Omega_{\rm \Lambda}$), and also include a term for spatial curvature ($\Omega_{\rm K} = 1-\Omega = 1 - \Omega_{\rm M} - \Omega_{\rm R} - \Omega_{\rm \Lambda}$), then we can express $H(z)$ \citep{weinberg2008}:
\begin{equation}
H(z) = H_{\rm 0} E(z)\,,
\end{equation}

\noindent where
\begin{equation}
E(z) = \sqrt{\Omega_{\rm R}(1+z)^{4} + \Omega_{\rm M}(1+z)^{3} + \Omega_{\rm K}(1+z)^{2} + \Omega_{\rm \Lambda} }\,.
\label{eqn:Ez}
\end{equation}

\noindent Note that the $\Omega_{i}$ values here correspond to those at $z=0$, and that the above expression does not include any redshift-dependent factors alongside $\Omega_{\rm \Lambda}$.  If desired, a model can also be included for a varying vacuum energy density, as discussed in Section~\ref{sec:hifi}.  

The dependence of a given quantity on the Hubble constant is often explicitly stated alongside the physical units of a measurement through the use of the dimensionless Hubble constant (`little h'):
\begin{eqnarray}
h \equiv \frac{H_{\rm 0}}{100\, \mathrm{km\, s^{-1} Mpc^{-1}}} \,.
\end{eqnarray}

The various uses of this quantity and its pitfalls are well-described in \citet{croton2013}.  In this paper, and to further aid the comparison of values contained in the historical literature, we give all Hubble Constant dependencies in terms of $h_{\rm C}$, allowing for the practice sometimes used of specifying the precise value of $H_{\rm 0}$ used in the little $h$ nomenclature: 
\begin{eqnarray}
h_{\rm C} \equiv  \frac{H_{\rm 0}}{C\,\, \mathrm{km\, s^{-1} Mpc^{-1}}} \,.
\end{eqnarray}

\noindent An explanation of how to convert values from one value of the Hubble constant to another with a specified $h_{\rm C}$ dependence is given in Appendix A.

\section{Cosmological Distances}
\label{sec:distances}

There are a number of cosmological distances relevant to the calculation and understanding of HI quantities, most notably including the line-of-sight comoving distance to a galaxy ($D_{\rm C}$), the transverse comoving distance ($D_{\rm M}$), the luminosity distance to a galaxy ($D_{\rm L}$), and the angular diameter distance ($D_{\rm A}$).  Following \citet{hogg1999}, the comoving distance $D_{\rm C}$ is given by:
\begin{eqnarray}
D_{\rm C}(z) = \frac{c}{H_{\rm 0}}\int_{0}^{z} E^{-1}(z')dz'\,,
\label{eqn:dc}
\end{eqnarray}

\noindent where $E(z)$ is as expressed in Equation~\ref{eqn:Ez}. As before, this can also be modified to include a model for a varying vacuum energy density if desired.  From $D_{\rm C}$ we can then express the transverse comoving distance $D_{\rm M}$:

\begin{equation}
 D_{\rm M}(z)= 
    \begin{cases}
    \frac{c}{H_{\rm 0}\sqrt{\Omega_{\rm K}}} {\rm sinh} \left( \frac{H_{\rm 0}\sqrt{\Omega_{\rm K}}}{c} D_{\rm C}\right) & \text{if } \Omega_{\rm K} > 0\,, \\
    D_{\rm C} & \text{if } \Omega_{\rm K} = 0\,,\\
    \frac{c}{H_{\rm 0}\sqrt{|\Omega_{\rm K}|}} {\rm sin} \left( \frac{H_{\rm 0}\sqrt{|\Omega_{\rm K}|}}{c} D_{\rm C}\right) & \text{if } \Omega_{\rm K} < 0\,,
    \end{cases}
\end{equation}

\noindent and finally the luminosity distance $D_{\rm L}$ and angular diameter distance $D_{\rm A}$:
 
\begin{eqnarray}
D_{\rm L}(z)  &=& (1+z)D_{\rm M}(z)\,, \\
D_{\rm A}(z) &=& D_{\rm M}(z)/(1+z)\,.
\end{eqnarray}

\noindent As these distances are all inversely proportional to the chosen value of the Hubble constant, this dependence can be included explicitly via little $h$ in the units, e.g. $h_{\rm C}^{-1}$ Mpc.  

\section{Line-of-Sight Velocity}

The various different definitions of velocity offer a significant potential source of confusion, particularly given the historical use of velocity in HI spectral line studies to describe both rest frame motions (i.e. the motions of objects {\it through} space), $V_{\rm pec}$, as well as a proxy for redshift caused by the expansion {\it of} space, $V_{\rm cos}$.  

Beginning with the full relativistic expressions for these two velocities, for $V_{\rm pec}$ we have the special relativity expression \citep{einstein1905, davis2003}:

\begin{eqnarray}
V_{\rm pec}(z_{\rm pec}) &=& V_{\rm SR}(z_{\rm pec}) \,, \\
	&=& c\, \frac{ \nu_{\rm rest}^2 - \nu_{\rm obs}^2 }{ \nu_{\rm rest}^2 + \nu_{\rm obs}^2 } \,, \\
         &=& c\, \frac{(1+z_{\rm pec})^2 - 1}{ (1+z_{\rm pec})^2 + 1 }  \,,
\end{eqnarray}

\noindent where these equations assume that motion is {\em purely} along the line of sight relative to the observer (see \citeauthor{einstein1905} for the relevant equations where the source has a transverse velocity component), and that the observer is in the same inertial reference frame as the source.  

In comparison, for $V_{\rm cos}$ we have the general relativistic expression \citep{davis2003}:

\begin{eqnarray}
V_{\rm cos}(z_{\rm cos}, z_{\rm ref}) &=& V_{\rm GR}(z_{\rm cos}, z_{\rm ref}) \,, \\
	&=& c \frac{H(z_{\rm ref})}{1+z_{\rm ref}}  \int^{z_{\rm cos}}_{0} \frac{dz'_{\rm cos}}{H(z'_{\rm cos})} \,,
\end{eqnarray}

\noindent where $z_{\rm ref}$ is the redshift at which the velocity is to be evaluated. In practice, as we are observing the object at $z_{\rm cos}$ at the current epoch, $z_{\rm ref}=0$.  In this situation, the above equation simplifies to (and as used in CosmoCalc):

\begin{eqnarray}
V_{\rm cos}(z_{\rm cos}) &=& c \int^{z_{\rm cos}}_{0} E^{-1}(z') dz' \,.
\end{eqnarray}

\noindent Comparing the above expression with Equation~\ref{eqn:dc}, it can be seen that $V_{\rm cos}$ then is the velocity that restores a basic Hubble Law relation $V_{\rm cos}(z) = D_{\rm C}(z)/H_{\rm 0}$ for all redshifts.

For small redshifts ($z<0.1$), both the special relativistic and general relativistic formulae can be approximated by the `optical' velocity convention (to distinguish this from the now deprecated `radio' velocity alternative):
\begin{equation}
V_{\rm opt}(z)=cz \,.
\end{equation}

\noindent As a general rule to avoid confusion, the use of velocities is best restricted to describe source {\em rest frame} motions (e.g. galaxy rotation, peculiar velocities), and not as a proxy for {\em observed frame} quantities (e.g. cosmological redshift, distance, or observed frame frequency width).  To separate these two potential uses of velocity in the rest of the paper, we refer to $V_{\rm rest}$ ($\equiv cz_{\rm pec}$, as in general for HI, $V_{\rm pec}\ll c$ and so this is a good approximation of $V_{\rm SR}$) and for the non-recommended observed frame velocity we refer to $V_{\rm obs}$ ($\equiv cz_{\rm obs}$, which while only an accurate approximation of $V_{\rm GR}$ at low redshift, has the advantage that it is readily invertible to obtain the source redshift if desired, and is the form that has traditionally been used in the literature).  For reference, a comparison plot showing the differences between $V_{\rm SR}(z)$, $V_{\rm GR}(z)$ and $V_{\rm opt}(z)$ can be found in \citet{davis2001}.

\section{Emission Profile Width}

Along with redshift, another key parameter that can be measured from HI profiles is their frequency width, which provides a line-of-sight measure of velocity differences in the source material, predominantly caused by galaxy rotation.  In a similar vein to the discussion of the previous section, to avoid confusion it is recommended that widths in the source rest frame be specified in terms of velocity, $\Delta V_{\rm rest}$, while those in the observed frame be given in frequency, $\Delta \nu_{\rm obs}$.  The relation between the two can be derived from Equation~\ref{eqn:zobs}.  Using the non-relativistic approximation for rest frame velocity width, $z_{\rm pec} = \pm V_{\rm pec}/c = \pm \Delta V_{\rm rest}/(2c)$, this gives:

\begin{eqnarray}
\Delta \nu_{\rm obs} &=& \frac{\nu_{\rm \HIsub}}{c(1+z_{\rm cos})\left(1 - \left(\frac{\Delta V_{\rm rest}}{2c} \right)^{2}\right)} \Delta V_{\rm rest} \label{eqn:nuobsVrest} \,.
\end{eqnarray}

\noindent Given the non-relativistic assumption $\Delta V_{\rm rest}\ll c$, and again assuming that any systematic peculiar velocity can be ignored ($z_{\rm obs} = z_{\rm cos} = z$), Equation~\ref{eqn:nuobsVrest} simplifies to the basic relations used in this paper to convert between $\Delta V_{\rm \rm rest}$ and $\Delta \nu_{\rm obs}$:

\begin{eqnarray}
\Delta V_{\rm rest} &\simeq& \frac{c(1+z)}{\nu_{\rm \HIsub}}\Delta \nu_{\rm obs} = \frac{c}{\nu_{\rm obs}}\Delta \nu_{\rm obs}\label{eqn:vrest} \,.
\end{eqnarray}

\noindent While discouraged, if you need to convert between observed frame and rest frame velocities (such as might need to be done when using some software which by default will measure observed frame widths in terms of optical velocity), or similarly between observed frame and rest frame frequency widths, this can be done using:

\begin{eqnarray}
\Delta V_{\rm rest} &=& \frac{1}{1+z}\Delta V_{\rm obs} \,, \\
\Delta \nu_{\rm obs} &=& \frac{1}{1+z}\Delta \nu_{\rm rest} \,.
\end{eqnarray} 

\noindent As a trivial extension, Equation~\ref{eqn:vrest} can also be used to determine the rest frame velocity resolution for a fixed observed frame channel width (setting $\Delta\nu_{\rm obs}=\Delta\nu_{\rm chan}$).

\section{Beam}

For a telescope with a normalised main beam sensitivity response that can be represented as a 2D elliptical gaussian (and ignoring position angle):

\begin{eqnarray}
P_{\rm n}(x,y) = e^{-\left(\frac{x^{2}}{2\sigma_{x}^{2}} + \frac{y^{2}}{2\sigma_{y}^{2}}\right)} \,,
 \end{eqnarray}
 
\noindent its `noise-equivalent field of view', or equivalent solid angle sampled at full sensitivity, is given by \citep[e.g.][]{kraus1986}:

\begin{eqnarray}
\Omega_{\rm bm} &=& \int^{\infty}_{-\infty}\int^{\infty}_{-\infty} P_{\rm n}(x,y)\, dx dy \,, \\
&=& 2\pi \sigma_{x}\sigma_{y} \,, \\
&=& \frac{\pi ab}{4\ln(2)} \,. 
\end{eqnarray}
 
\noindent where $a$ ($ = 2\sqrt{2\ln(2)}\sigma_{x}$) is the main beam angular major axis, and $b$ ($ = 2\sqrt{2\ln(2)} \sigma_{y}$) is the angular minor axis, both measured at the half power point.   

Note that the half power points $a,b \propto \lambda_{\rm obs} \propto (1+z)$, so for a synthesised beam with no frequency-dependent weighting, or a non-compound primary beam, the beam solid angle will vary as:

\begin{eqnarray}
\Omega_{\rm bm}(z) &= (1+z)^{2}\Omega_{\rm bm}(0) \label{eqn:beam_area_omega} \,.
\end{eqnarray}

\noindent In practice, the illumination of a telescope is frequency dependent, so the beam area may not scale exactly as $(1+z)^{2}$ as given above. A broadband feed will generally under-illuminate a dish at the highest frequencies, making the beam larger (and the efficiency lower).

\section{Flux Density}

The observed flux density, $S_{\nu}$, for a source with a rest frame luminosity density of  $L_{\nu_{\rm rest}} = L_{(1+z)\nu_{\rm obs}} = L_{(1+z)\nu}$ will be \citep{peacock1999}:

\begin{eqnarray}
S_{\nu} = (1+z)\frac{L_{(1+z)\nu}}{4\pi D_{L}^{2}} \,. 
\end{eqnarray}

\noindent The preferred unit for $S_{\nu}$ is Jy (= $10^{-26}\,{\rm W\,m^{-2}\, Hz^{-1}})$.  The maximum of $S_{\nu}$ for a given source is often referred to an object's `peak' flux density.  

As a side note, the intensity scale of HI images is usually presented in units of `Jy per beam', reflecting the fact that a telescope will measure any flux within its beam when pointed a source, i.e. that the specific intensity observed, $I_{\nu}^{\rm obs}$, is the real specific intensity at the location of the observer, $I_{\nu}$, convolved with the normalised telescope beam, $P_{\rm n}$. The flux density, $S_{\nu}$ of the source is then given by \citep{kraus1986}:

\begin{eqnarray}
S_{\nu} &=& \int_{\Omega_{\rm src}} I_{\nu}(\theta,\phi) \,{\rm d}\Omega  \,,\\ 
              &=& \frac{1}{\Omega_{\rm bm}}\int_{\Omega_{\rm src}} I_{\nu}(\theta,\phi) \ast P_{\rm n}(\theta,\phi) \,{\rm d}\Omega \,,\\
              &=& \frac{1}{\Omega_{\rm bm}}\int_{\Omega_{\rm src}} I_{\nu}^{\rm obs}(\theta,\phi)\,{\rm d}\Omega \,.
\end{eqnarray}

To recover the correct flux for an extended region, $\int_{\Omega_{\rm src}} I_{\nu}^{\rm obs}(\theta,\phi)\,{\rm d}\Omega$ is measured by summing values of the pixels in the source region (in Jy per beam) multiplied by the pixel area, from which $S_{\nu}$ can then recovered by dividing by the area of the telescope beam.  Or expressed alternatively, $S_{\nu}$ can be measured by simply converting the specific intensity values of the image to Jy per pixel, and then summing over the region of interest. 

Note that if the ratio of beam area to pixel area is constant as a function of frequency (i.e. both pixel area and beam area are scaling as $1+z$), then the correction factor for the extended source flux sum will be constant, but if not, e.g. the image cube has a fixed angular pixel sale, a varying correction will be required.

For a point source, while this approach could also be taken, it results in a sub-optimal signal-to-noise measurement compared to either just taking the value of the pixel centred on the source (although this has the potential to underestimate the source flux if the pixel is not exactly centred on the source), or weighting the pixel values by the telescope beam response (which avoids this problem, but increases the effective angular area over which the measurement is being made compared to the single pixel method, and as such can increase the impact of source confusion).

\section{Flux}

For a source at redshift z, with a rest frame total HI luminosity of $L$, its observed flux $S$ will be:

\begin{eqnarray}
S &=& \int S_{\nu} d\nu_{\rm obs} \,, \\ 
    &=& \int (1+z)\frac{L_{\nu_{\rm rest}}}{4\pi D_{L}^{2}} d\nu_{\rm obs} \,,\\
    &=& \int \frac{L_{\nu_{\rm rest}}}{4\pi D_{L}^{2}} d\nu_{\rm rest} \,, \\
\Rightarrow S &=& \frac{L}{4\pi D_{L}^{2}} \,.
\end{eqnarray}

\noindent The preferred unit for this quantity is Jy Hz (= $10^{-26}\,{\rm W\,m^{-2})}$.  In HI studies, reference is often made to a similar, but dimensionally different quantity defined as the integral of flux density as a function of velocity rather than frequency:

\begin{eqnarray}
S^{V} = \int S_{\nu} dV \,,
\end{eqnarray}

\noindent the units for which are Jy km s$^{-1}$ (= $10^{-26}\,{\rm W\,m^{-2} Hz^{-1}\, km\, s^{-1}}$).  This is a poor quantity for cosmological measurements as it introduces some uncertainty about exactly what velocity (or pseudo-velocity) has been used in its calculation, as discussed earlier.  It is preferable to use $S$ rather than $S^{V}$ wherever possible, and particularly so if referring to a measurement of an object's observed total flux given the dimensional difference of $S^{V}$ from that of a natural flux value.

The conversion between $S$ and $S^{V}$ depends on the velocity convention used.  If $S^{V}$ is measured in the optical observed frame:

\small\begin{eqnarray} 
\left( \frac{S^{V_{\rm obs}}}{\rm Jy\, km\, s^{-1}} \right) &=& \frac{c(1+z)^{2}}{\nu_{\HIsub}} \left( \frac{S}{\rm Jy\, Hz} \right) , \\
 & \simeq & 2.11\times10^{-4} \left(1+z\right)^{2} \left( \frac{S}{\rm Jy\, Hz} \right) .
\end{eqnarray}\normalsize

\noindent Alternatively, if the velocity used for the calculation of $S^{V}$ is a source rest frame velocity (for which $V_{\rm opt} \simeq V_{\rm SR}$ at the velocities relevant for galaxy rotation), then:

\small\begin{eqnarray} 
\left( \frac{S^{V_{\rm rest}}}{\rm Jy\, km\, s^{-1}} \right) &=& \frac{c(1+z)}{\nu_{\HIsub}} \left( \frac{S}{\rm Jy\, Hz} \right) , \\
&\simeq & 2.11\times10^{-4} \left(1+z\right) \left( \frac{S}{\rm Jy\, Hz} \right) .
\end{eqnarray}\normalsize

\section{Number of HI Atoms}

Taking $\frac{3}{4}$ of HI atoms to be in the upper hyperfine state, with a spontaneous emission rate of $A_{\HIsub}$, an emitted photon energy of $h\nu_{\HIsub}$, and an HI source with luminosity $L$ to be optically thin, the number of HI atoms, $\mathscr{N}_{\HIsub}$, will be given by:

\begin{eqnarray}
\mathscr{N}_{\HIsub} &=&  \frac{L}{\frac{3}{4} h\nu_{\HIsub} A_{\HIsub}} \,, \\
            &=&  \frac{16\pi D_{L}^{2}S}{3 h\nu_{\HIsub} A_{\HIsub}} \,,
\label{eqn:natoms}
\end{eqnarray}

\small
\begin{eqnarray}
\Rightarrow \left( \frac{\mathscr{N}_{\HIsub}}{h_{\rm C}^{-2}}\right) \simeq 5.91\times 10^{58} \left( \frac{D_{L}}{h_{\rm C}^{{-1}}\rm Mpc} \right)^{2} \left( \frac{S}{\rm Jy\, Hz} \right) .
\end{eqnarray}
\normalsize

\section{HI Mass}

Using the above relation, the HI mass of a source is trivially given by (caveat corrections that need to be applied if the source is not optically thin):

\begin{eqnarray} 
M_{\HIsub} &=& \mathscr{N}_{\HIsub}m_{\rm H} \,,
\end{eqnarray}

\small\begin{eqnarray}
\Rightarrow \left( \frac{M_{\HIsub}}{h_{\rm C}^{-2}M_{\odot}} \right) &\simeq& 49.7 \, \left( \frac{D_L}{h_{\rm C}^{-1}\rm Mpc} \right)^{2 }\left( \frac{S}{\rm Jy\, Hz} \right) . \label{eqn:HImass} 
\end{eqnarray}\normalsize

\noindent Which compares to the relation expressed in terms of traditional observed frame velocity integrated flux:

\small\begin{eqnarray} 
\left( \frac{M_{\HIsub}}{h_{\rm C}^{-2}M_{\odot}} \right) \simeq \frac{2.35 \times 10^{5}}{(1+z)^{2}} \left( \frac{D_L}{h_{\rm C}^{-1}\rm Mpc} \right)^{2 }\left( \frac{S^{V_{\rm obs}}}{\rm Jy\, km s^{-1}} \right) .\label{eqn:HImassSvobs} 
\end{eqnarray}\normalsize

\noindent Or rest frame velocity integrated flux:

\small\begin{eqnarray} 
\left( \frac{M_{\HIsub}}{h_{\rm C}^{-2}M_{\odot}} \right) \simeq \frac{2.35 \times 10^{5}}{1+z} \left( \frac{D_L}{h_{\rm C}^{-1}\rm Mpc} \right)^{2 }\left( \frac{S^{V_{\rm rest}}}{\rm Jy\, km s^{-1}} \right) .\label{eqn:HImassSvrest} 
\end{eqnarray}\normalsize

\section{HI Mass Function}

The HI mass function, $\Theta(M_{\HIsub})$, gives the number of sources per unit volume as a function of HI mass, i.e. the volume density of sources, $n$, having masses between M1 and M2 will be given by:

\begin{eqnarray}
n = \int_{\hat{M}1}^{\hat{M}2} \Theta(\hat{M}_{\HIsub})d\hat{M}_{\HIsub} \,,
\end{eqnarray}

\noindent where for convenience in the equations that follow we have expressed this in terms of dimensionless mass $\hat{M}_{\HIsub}\equiv M_{\HIsub}/M_{\HIsub}^{*}$, $M_{\HIsub}^{*}$ being one of the Schechter function parameters traditionally used to parametrize the mass function.  Clearly the mass function could also be expressed without this change of variable.  The Schechter function form is given by \citep{schechter1976,zwaan2005}:
 
\begin{eqnarray}
\Theta(\hat{M}_{\HIsub})d\hat{M}_{\HIsub} = \theta^{*}  \hat{M}_{\HIsub}^{\alpha} e^{-\hat{M}_{\HIsub}} d\hat{M}_{\HIsub}  \label{eqn:schechter}\,, \\
\end{eqnarray}

\noindent where $\theta^{*}$ provides the overall normalisation, $\alpha$ gives the slope of the low mass power-law, and $M_{\HIsub}^{*}$ gives the characteristic turnover mass, above which the number of sources exponentially declines.  Alternatively, the number of sources is often expressed as an integral over $\log(\hat{M}_{\HIsub})$ rather than $\hat{M}_{\HIsub}$, ie:

\begin{eqnarray}
n = \int_{\log(\hat{M}1)}^{log(\hat{M}2)} \Phi(\hat{M}_{\HIsub})d \log \hat{M}_{\HIsub} \,, \label{eqn:schechterlog1}
\end{eqnarray}

\noindent which gives \citep{springob2005,martin2010}:  

\begin{dmath}
\Phi(\hat{M}_{\HIsub}) d\log \hat{M}_{\HIsub} = \phi^{*} \ln(10) \hat{M}_{\HIsub}^{\alpha+1}  e^{-\hat{M}_{\HIsub}} d\log \hat{M}_{\HIsub} \,. \label{eqn:schechterlog2}
\end{dmath}

\noindent In practice, the HI mass function is normally presented as $\log(\Phi)$ vs. $\log(\hat{M}_{\HIsub})$.  In this space, the functional form of the Schechter function is given by:

\begin{dmath}
\log \Phi =  \log(\phi^{*}\ln10) + (1+\alpha)\log\hat{M}_{\HIsub} - \hat{M}_{\HIsub} \log(e) \,.
\end{dmath}

\noindent For evolutionary studies, the mass function is best expressed in comoving coordinates to remove normalisation changes caused purely by the expansion of the Universe.  

Note also that the Schechter form of the HI mass function is not consistently defined in the literature, and so you may need to pay attention to the particular definition used when comparing results.

\section{Cosmological Mass Density}

The cosmological mass density of HI can be calculated by carrying out the mass-weighted integral over the HI mass function:

\begin{eqnarray}
\rho_{\HIsub} &=& \int_{0}^{\infty} M_{\HIsub} \Theta(M_{\HIsub})dM_{\HIsub} \,.
\end{eqnarray}

\noindent This can either be simply summed directly from the measured HI mass function data points (providing a sufficient mass range is spanned and making sure to correct $\Theta$ values to those appropriate for the $\log M_{\HIsub}$ binwidths actually used), or through the following analytic solution expressed in terms of the fitted Schechter parameters and the complete Gamma function, $\Gamma$ \citep{zwaan2003}:

\begin{align}
\left(\frac{\rho_{\HIsub}}{h_{\rm C}\, M_{\odot}\, {\rm Mpc}^{-3}}\right) =& \nonumber \\
\left(\frac{\theta^{*}}{h_{\rm C}^{3}\,{\rm Mpc}^{-3}}\right)& \Gamma(2+\alpha)\left( \frac{M_{\HIsub}^{*}}{h_{\rm C}^{-2}M_{\odot}}\right) .
\end{align}\\

\noindent For an HI mass function measured using comoving volume, this will similarly yield a cosmic HI density in comoving coordinates, and so $\rho_{\HIsub}$ should remain constant with redshift/lookback time if no evolution is occurring aside from the expansion of the Universe.

The cosmic mass density of HI can also be expressed as a fraction of the critical mass density:

\begin{eqnarray}
\Omega_{\HIsub}(z) = \frac{\rho_{\HIsub}(z)}{\rho_{\rm crit}(z)} \,,
\end{eqnarray}

\noindent where in proper (physical) coordinates $\rho_{\rm crit}(z)$ is given by:
\begin{eqnarray}
\rho_{\rm crit}^{p}(z) = \frac{3H^{2}(z)}{8\pi G} \,,
\end{eqnarray}

\noindent or in comoving coordinates:
\begin{eqnarray}
\rho_{\rm crit}^{c}(z) = \frac{1}{(1+z)^{3}}\frac{3H^{2}(z)}{8\pi G} \,.
\end{eqnarray}
\noindent Evaluating these expressions for $\Omega_{\HIsub}$ gives:
\begin{dmath}[breakdepth={1},style={\small}]
\left( \frac{\Omega_{\HIsub}(z)}{h_{\rm C}^{-1}} \right) = \left( \frac{3.60 \times 10^{-8} \times H^{-2}(z)}{h_{\rm C}^{-2} M_{\odot}^{-1}{\rm Mpc}^{3}} \right) \times \left( \frac{\rho_{\HIsub}^{p}(z)}{h_{\rm C} M_{\odot} {\rm Mpc}^{-3}} \right) ,
\end{dmath}
\noindent or:
\begin{dmath}[breakdepth={1},style={\small}]
\left( \frac{\Omega_{\HIsub}(z)}{h_{\rm C}^{-1}} \right) = \frac{1}{(1+z)^{3}}\left( \frac{3.60 \times 10^{-8} \times H^{-2}(z)}{h_{\rm C}^{-2} M_{\odot}^{-1}{\rm Mpc}^{3}} \right) \left( \frac{\rho_{\HIsub}^{c}(z)}{h_{\rm C} M_{\odot} {\rm Mpc}^{-3}} \right) .
\end{dmath}

\noindent Note that for an expanding Universe, if there exist contributors to the total energy density other than matter, $\rho_{\rm crit}^{c}$ will evolve with redshift along with $\rho_{\rm crit}^{p}$.  As such, $\Omega_{\HIsub}$ will also change, even if the comoving density of HI does not.  To remove this effect, the practice is often used of replacing $\rho_{\rm crit}(z)$ with $\rho_{\rm crit}(z=0)$ in the above expressions, ie:
\begin{eqnarray}
\Omega_{\HIsub}(z) = \frac{\rho_{\HIsub}(z)}{\rho_{\rm crit}(z=0)} \,,
\end{eqnarray}
\noindent which also offers the convenience of $\rho_{\rm crit}^{p}(z=0) = \rho_{\rm crit}^{c}(z=0)$, but be aware of the difference between this and the above derivations, and be sure to specify which definition of $\Omega_{\HIsub}(z)$ is being used.

\section{Brightness Temperature}

The brightness temperature corresponding to an observed flux density $S_{\nu}$, measured with a telescope of main beam solid angle $\Omega_{\rm bm}$, is the blackbody temperature an extended object would need to have to produce the observed flux in the Rayleigh-Jeans limit \citep[$h\nu \ll kT$; ][]{wilson2009}.  From the Rayleigh-Jeans law, the luminosity emitted by a blackbody per unit area into a unit solid angle, $B_{\nu}$ is given by:

\begin{eqnarray}
B_{\nu} = \frac{2 k \nu^{2} T}{c^{2}} \,,
\end{eqnarray}

\noindent where $k$ is Boltzmann's constant.  In a Euclidean geometry, $I_{\nu} = B_{\nu}$ (surface brightness conservation; $I_{\nu}$ is the received specific intensity), giving the traditional local-Universe relation for HI brightness temperature:

\begin{eqnarray}
B_{\nu} &=& I_{\nu} = \frac{S_{\nu}}{\Omega_{\rm bm}}  =  \frac{2 k \nu_{\HIsub}^{2} T}{c^{2}} \,, \\
\Rightarrow T_{\rm B} &=&  \frac{c^{2} S_{\nu}}{2k\nu_{\HIsub}^{2}\Omega_{\rm bm}} \,.
\end{eqnarray}

\noindent However, in a relativistic Universe, we have \citep{peacock1999}:

\begin{eqnarray}
I_{\nu} &=& \frac{B_{\nu(1+z)}}{(1+z)^{3}} = \frac{B_{\nu_{\rm rest}}}{(1+z)^{3}} \,,
\end{eqnarray}

\noindent giving:

\begin{eqnarray}
I_{\nu} = \frac{S_{\nu}}{\Omega_{\rm bm}} = \frac{2 k \nu_{\HIsub}^{2}T}{c^{2}(1+z)^{3}} \,, \\
\Rightarrow  T_{\rm B} = (1+z)^{3}\frac{c^{2}S_{\nu}}{2 k \nu_{\HIsub}^{2}\Omega_{\rm bm}} \,. \label{eqn:TB}
\end{eqnarray}

\noindent Evaluating the constants gives:

\begin{eqnarray}
\left( \frac{T_{\rm B}}{K} \right) = 6.86 \times 10^{5} \left( 1+ z \right)^{3} \left( \frac{S_{\nu}}{\rm Jy} \right)\left( \frac{\Omega_{\rm bm}}{\rm arcsec^{2}} \right)^{-1} \!\!\!\!\!\!,
\end{eqnarray}

\noindent or expressed alternatively as a function of the beam angular major and minor axes $a$ and $b$:

\begin{eqnarray}
\left( \frac{T_{\rm B}}{K} \right) = 6.06 \times 10^{5} \left( 1+ z \right)^{3} \left( \frac{S_{\nu}}{\rm Jy} \right)\left( \frac{ab}{\rm arcsec^{2}} \right)^{-1} \!\!\!\!\!\!.
\end{eqnarray}

\section{Column Density}

The HI column density gives the number of atoms per unit area along the line of sight through an astronomical object.  The column density $N_{\HIsub}$ for a flux $S$ measured over solid angle $\Omega$ is given by (using Equation~\ref{eqn:natoms} for the number of atoms, and the small angle approximation to calculate the source area; material is again assumed to be optically thin):

\begin{eqnarray}
N_{\HIsub} &=& \frac{\mathscr{N}_{\HIsub}}{\rm area} = \left( \frac{16\pi D_{L}^{2}S}{3 h\nu_{\HIsub} A_{\HIsub}} \right)\left( \frac{1}{D_{A}^{2}\Omega_{\rm bm}} \right)\,,\\
                 &=& \frac{16\pi }{3 h\nu_{\HIsub} A_{\HIsub}}  (1+z)^{4}  \frac{S}{\Omega_{\rm bm}}\,,\label{eqn:nHI}
\end{eqnarray}

\small\begin{dmath}
\Rightarrow \left(\frac{N_{\HIsub}}{\rm cm^{-2}}\right) = 2.64 \times 10^{20} (1+z)^{4}\left(\frac{S}{\rm JyHz} \right) \left( \frac{\Omega_{\rm bm}}{\rm arcsec^{2}}\right)^{-1} \!\!\!\!\!\!.
\end{dmath}\normalsize

\noindent Or in terms of the beam angular major and minor axes $a$ and $b$:

\begin{eqnarray}
N_{\HIsub} &=& \left( \frac{16\pi D_{L}^{2}S}{3 h\nu_{\HIsub} A_{\HIsub}} \right)\left( \frac{4\ln(2)}{D_{A}^{2}\pi ab } \right) ,
\end{eqnarray}

\small\begin{dmath}
\Rightarrow \left(\frac{N_{\HIsub}}{\rm cm^{-2}}\right) = 2.33 \times 10^{20} (1+z)^{4}\left(\frac{S}{\rm JyHz} \right) \left( \frac{ab}{\rm arcsec^{2}}\right)^{-1} \!\!\!\!\!\!.
\end{dmath}\normalsize

\noindent For comparison, the expressions in terms of observed frame velocity integrated flux are:

\small\begin{eqnarray}
\left(\frac{N_{\HIsub}}{\rm cm^{-2}}\right)  &=& 1.25 \times 10^{24} (1+z)^{2} \\ \nonumber && \times \left( \frac{S^{V_{\rm obs}}}{\rm Jy\, km s^{-1}} \right) \left( \frac{\Omega_{\rm bm}}{\rm arcsec^{2}}\right)^{-1} \!\!\!\!\!\!, \\
\left(\frac{N_{\HIsub}}{\rm cm^{-2}}\right)  &=& 1.10 \times 10^{24} (1+z)^{2} \\ \nonumber && \times \left( \frac{S^{V_{\rm obs}}}{\rm Jy\, km s^{-1}} \right) \left( \frac{ab}{\rm arcsec^{2}}\right)^{-1} \!\!\!\!\!\!.
\end{eqnarray}\normalsize

\noindent And for rest frame velocity integrated flux:

\small\begin{eqnarray}
\left(\frac{N_{\HIsub}}{\rm cm^{-2}}\right) &=& 1.25 \times 10^{24} (1+z)^{3} \\ \nonumber && \times \left( \frac{S^{V_{\rm rest}}}{\rm Jy\, km s^{-1}} \right) \left( \frac{\Omega_{\rm bm}}{\rm arcsec^{2}}\right)^{-1} \!\!\!\!\!\!,\\
\left(\frac{N_{\HIsub}}{\rm cm^{-2}}\right)  &=& 1.10 \times 10^{24} (1+z)^{3} \\ \nonumber && \times \left( \frac{S^{V_{\rm rest}}}{\rm Jy\, km s^{-1}} \right) \left( \frac{ab}{\rm arcsec^{2}}\right)^{-1} \!\!\!\!\!\!.
\end{eqnarray}\normalsize

\noindent The above column density expressions can be readily expressed as HI surface mass densities by multiplying $m_{\rm H}$ and converting to the desired units, eg.:

\small\begin{eqnarray}
\left(\frac{\Sigma _{\rm HI}}{\rm M_{\odot}\, pc^{-2}}\right) &=&  8.01 \times 10^{-21}\left(\frac{N_{\HIsub}}{\rm cm^{-2}}\right) , \\
 &=&  2.12 \,\, (1+z)^{4}\left(\frac{S}{\rm JyHz} \right) \left( \frac{\Omega_{\rm bm}}{\rm arcsec^{2}}\right)^{-1} \!\!\!\!\!\!, \\
 &=&  1.00 \times 10^{4}  (1+z)^{2} \\ \nonumber && \times \left( \frac{S^{V_{\rm obs}}}{\rm Jy\, km s^{-1}} \right) \left( \frac{\Omega_{\rm bm}}{\rm arcsec^{2}}\right)^{-1} \!\!\!\!\!\!, \\
 &=&  1.00 \times 10^{4}  (1+z)^{3} \\ \nonumber && \times \left( \frac{S^{V_{\rm rest}}}{\rm Jy\, km s^{-1}} \right) \left( \frac{\Omega_{\rm bm}}{\rm arcsec^{2}}\right)^{-1} \!\!\!\!\!\!. 
\end{eqnarray}\normalsize

\noindent Column density can also be expressed as a function of brightness temperature, $T_{\rm B}$.  Following \citet{wilson2009,giovanelli1988,brinks1990}, for a cloud with spin temperature $T_{\rm S}$ and optical depth $\tau$ embedded in a radiation field having brightness temperature $T_{\rm R}$, its brightness temperature, $T_{\rm B}$, will be given by:

\begin{eqnarray}
T_{\rm B} = T_{\rm R} e^{-\tau_{\nu}} + T_{\rm S}(1-e^{-\tau_{\nu}}) \,.
\end{eqnarray}

\noindent If the cloud is optically thin ($\tau \ll1$), and assuming that background radiation field has negligible impact, then:

\begin{eqnarray}
T_{\rm B} = \tau_{\nu} T_{\rm S} \,. \label{eqn:TbtauTs}
\end{eqnarray}

\noindent The optical depth per unit frequency can be related to the density of hydrogen atoms per unit length along the line of sight:

\begin{eqnarray}
d\tau_{\nu} = - \frac{3c^{2}}{32\pi \nu_{\HIsub}}A_{\HIsub} n_{\HIsub}\frac{h}{kT_{\rm s}}\phi(\nu) dl \,,
\end{eqnarray}

\noindent where $n_{\HIsub}$ refers to the volume density of hydrogen atoms, and $\phi$ is the line shape function reflecting the natural line width ($\int \phi(\nu) d\nu=1$). Integrating this over $\nu$ and $s$ gives:

\begin{eqnarray}
\int n_{\HIsub} dl = \frac{32\pi k \nu_{\HIsub}}{3A_{\HIsub}hc^{2}} T_{\rm s} \int \tau_{\nu}d\nu \,.
\end{eqnarray}

\noindent The left-hand side of this expression is the desired HI column density $N_{\HIsub} = \int n_{\HIsub} dl$, while the right-hand side can now be written as a function of brightness temperature using Equation~\ref{eqn:TbtauTs}:

\begin{eqnarray}
N_{\HIsub} &=& \frac{32\pi k \nu_{\HIsub}}{3A_{\HIsub}hc^{2}} \int T_{\rm B} d\nu_{\rm rest} \,, \label{eqn:nHI_TB}\\
\Rightarrow \left( \frac{N_{\HIsub}}{\rm cm^{-2}} \right) &=& 3.85 \times 10^{20} \int \left( \frac{T_{\rm B}}{\rm K} \right) \left( \frac{d\nu_{\rm rest}}{\rm MHz} \right) \!. \label{eqn:nHI_TB_val}
\end{eqnarray}

\noindent Note that the frequency integral here is in the source rest frame.  As both $N_{\HIsub}$ and $T_{\rm B}$ are also rest frame quantities, there are no $(1+z)$ factors in the above expression.  Alternatively, Equation~\ref{eqn:nHI_TB} can also expressed as an integral over rest frame velocity:

\begin{eqnarray}
N_{\HIsub} &=& \frac{32\pi k \nu_{\HIsub}^{2}}{3A_{\HIsub}hc^{3}} \int T_{\rm B} dV_{\rm rest} \,, \\
\Rightarrow \left( \frac{N_{\HIsub}}{\rm cm^{-2}} \right) &=& 1.82 \times 10^{18} \int \left( \frac{T_{\rm B}}{\rm K}\right) \left( \frac{dV_{\rm rest}}{\rm km s^{-1}} \right) \!. \label{eqn:nHI_TB_Vrest_val}
\end{eqnarray}

\noindent To recover the previous expression for HI column density (Equation~\ref{eqn:nHI}), Equations~\ref{eqn:TB}~and~\ref{eqn:nHI_TB} give: 

\begin{eqnarray}
N_{\HIsub} &=& \frac{32\pi k \nu_{\HIsub}}{3A_{\HIsub}hc^{2}} \int (1+z)^{3}\frac{c^{2}S_{\nu}}{2 k \nu_{\HIsub}^{2}\Omega_{\rm bm}} d\nu_{\rm rest}\,, \\
 &=& \frac{16\pi }{3h\nu_{\HIsub}A_{\HIsub}}(1+z)^{4}\frac{1}{\Omega_{\rm bm}} \int S_{\nu} d\nu_{\rm obs} \,, \\
 &=& \frac{16\pi }{3h\nu_{\HIsub}A_{\HIsub}}(1+z)^{4}\frac{S}{\Omega_{\rm bm}} \hspace{1mm}\mbox{, as before.}
 \end{eqnarray}

\section{Flux and Flux Density Sensitivity}

\noindent If not known observationally, the flux density sensitivity of a single-pointing, naturally-weighted, Stokes-I, non-primary beam corrected interferometric image can be estimated using \citep{taylor2008,kraus1986,obreschkow2011}:

\begin{eqnarray}
\sigma_{S_{\nu}} &=& \frac{SEFD}{\eta_{\rm s}\sqrt{2 N(N-1)\Delta t \Delta \nu}} \,, \\
                            &=& \frac{k T_{\rm sys}}{\eta_{\rm s}A_{\rm e}}  \sqrt{\frac{2}{N(N-1)\Delta t \Delta \nu}} \,, \label{eqn:sigsnu}
\end{eqnarray}

\noindent where SEFD is the System Equivalent Flux Density of an antenna in the array ($=2 k T_{\rm sys}/A_{\rm e}$; assumed to be the same for all elements in the above relation), $\Delta \nu$ is the frequency width of interest, $\Delta t$ is the integration time, $T_{\rm sys}$ is the system temperature, $N$ is the number of antennae, $A_{\rm e}$ is the effective area of each antenna, $\eta_{\rm s}$ is the system efficiency, and $k$ is Boltzmann's constant.  It should be noted that in practice a number of factors may reduce the achieved observational sensitivity from the theoretical value above, such as the application of weighting and tapering.  Evaluating the constants to give the sensitivity in Jy:  
\small\begin{dmath}
\left(\frac{\sigma_{S_{\nu}}}{\rm Jy}\right) = \frac{1952.5}{\eta_{\rm s}\sqrt{N(N-1)}} \times \left(\frac{A_{\rm e}/T_{\rm sys}}{\rm m^{2} K^{-1}}\right)^{-1} \left(\frac{\Delta t}{\rm s}\right)^{-\frac{1}{2}} \left(\frac{\Delta \nu}{\rm Hz}\right)^{-\frac{1}{2}} \,.
\end{dmath}\normalsize

\noindent The corresponding flux sensitivity scaling is:

\begin{eqnarray}
\sigma_{S} &=& \Delta \nu\,  \sigma_{S_{\nu}} \,, \\
		  &=& \frac{SEFD}{\eta_{\rm s}}  \sqrt{\frac{\Delta \nu}{2 N(N-1)\Delta t}} \,, \\
                   &=& \frac{k T_{\rm sys}}{\eta_{\rm s}A_{\rm e}}  \sqrt{\frac{2\Delta \nu}{N(N-1)\Delta t}}  \,. \label{eqn:sigs}
\end{eqnarray}

\noindent Or:

\small\begin{dmath}
\left(\frac{\sigma_{S}}{\rm JyHz}\right) = \frac{1952.5}{\eta_{\rm s}\sqrt{N(N-1)}} \times \left(\frac{A_{\rm e}/T_{\rm sys}}{\rm m^{2} K^{-1}}\right)^{-1} \left(\frac{\Delta t}{\rm s}\right)^{-\frac{1}{2}} \left(\frac{\Delta \nu}{\rm Hz}\right)^{\frac{1}{2}} \,.
\end{dmath}\normalsize

Observationally, if the flux density sensitivity averaged over channel interval $\Delta \nu_{\rm chan}$ is known, $\sigma_{S_{\nu_{\rm chan}}}$, the flux and flux density sensitivities over interval $\Delta \nu$ can be simply calculated using:

\begin{eqnarray}
\sigma_{S_{\nu}} &=&  \sigma_{S_{\nu_{\rm chan}}}\sqrt{\frac{\Delta \nu_{\rm chan}}{\Delta \nu}} \,, \label{eqn:noiseSnu}\\
\sigma_{S} &=&  \sigma_{S_{\nu_{\rm chan}}}\sqrt{\Delta \nu_{\rm chan} \Delta \nu} \,. \label{eqn:noiseS}
\end{eqnarray}

As observed frame quantities, these have no inherent scaling as a function of redshift, however telescope performance may varying significantly as a function of frequency, often encoded as variations in the $A_{\rm e}/T_{\rm sys}$ term of Equations~\ref{eqn:sigsnu}~\&~\ref{eqn:sigs} (and system efficiency $\eta_{\rm s}$ if desired), introducing a redshift scaling of:

\begin{eqnarray}
\sigma_{S_{\nu}}(z) = f(z)\sigma_{S_{\nu}}(0) \,, \\  
\sigma_{S}(z) = f(z)\sigma_{S}(0) \,,
\end{eqnarray}

\noindent where:
\begin{eqnarray}
f(z) = \frac{\frac{\eta_{\rm s}A_{\rm e}}{T_{\rm sys}}(z=0) }{ \frac{\eta_{\rm s}A_{\rm e}}{T_{\rm sys}}(z) }\,. \label{eqn:ATscaling}
 \end{eqnarray}

\noindent As will become apparent in the next sections concerning brightness temperature and column density sensitivity, it can also be useful to understand how these relations vary as a function of rest frame velocity interval $\Delta V_{\rm rest}$.  Substituting Equation~\ref{eqn:vrest} into Equations~\ref{eqn:noiseSnu}~\&~\ref{eqn:noiseS} gives:

\begin{eqnarray}
\sigma_{S_{\nu}} =  \sigma_{S_{\nu_{\rm chan}}}\sqrt{\frac{ c(1+z)\Delta\nu_{\rm chan}}{\nu_{\HIsub}\Delta V_{\rm rest}}} \,, \label{eqn:noiseSnuVrest}\\
\sigma_{S} =  \sigma_{S_{\nu_{\rm chan}}}\sqrt{\frac{\nu_{\HIsub}\Delta V_{\rm rest}\Delta\nu_{\rm chan}}{c(1+z)}} \,, \label{eqn:noiseSVrest}
\end{eqnarray} 

\noindent and a redshift scaling of (with $A_{\rm e}/T_{\rm sys}$ variations):

\begin{eqnarray}
\sigma_{S_{\nu}}^{\Delta V_{\rm rest}^{\rm const}} (z) = f(z)(1+z)^{{\frac{1}{2}}} \sigma_{S_{\nu}}^{\Delta V_{\rm rest}^{\rm const}}(0) \,, \\
\sigma_{S}^{\Delta V_{\rm rest}^{\rm const}} (z) = f(z)(1+z)^{{-\frac{1}{2}}} \sigma_{S}^{\Delta V_{\rm rest}^{\rm const}}(0) \,. 
\end{eqnarray}

\subsubsection*{Widefield Single-Beam Mosaic:}

\noindent For a widefield mosaic in which many overlapping primary beam pointings have been used to image a large area, an additional factor needs to be considered.  For a fixed pointing pattern, the primary beam area for each pointing will also increase as $(1+z)^2$, which will in turn lead to the noise decreasing as $(1+z)^{-1}$ \citep{abdalla2005}, giving:

\begin{eqnarray}
\bar{\sigma}_{S_{\nu}}(z) &=& f(z)(1+z)^{-1} \bar{\sigma}_{S_{\nu}}(0) \,, \\
\bar{\sigma}_{S} (z) &=& f(z)(1+z)^{-1} \bar{\sigma}_{S}(0) \,, \\
\bar{\sigma}_{S_{\nu}}^{\Delta V_{\rm rest}^{\rm const}} (z) &=& f(z)(1+z)^{{-\frac{1}{2}}} \bar{\sigma}_{S}^{\Delta V_{\rm rest}}(0) \,, \\
\bar{\sigma}_{S}^{\Delta V_{\rm rest}^{\rm const}} (z) &=& f(z)(1+z)^{{-\frac{3}{2}}} \bar{\sigma}_{S}^{\Delta V_{\rm rest}}(0) \,.
\end{eqnarray}

\subsubsection*{Phased Array Feed Observations:}

\noindent For a phased array feed (PAF), the noise-equivalent field of view (${\rm FoV\!}_{\rm NE}$) will not increase as $(1+z)^{2}$ as in the single beam case above, due to correlated noise effects between adjacent formed beams. Indeed, once the central part of the field is fully sampled, ${\rm FoV\!}_{\rm NE}$ will not continue to rise with redshift apart from some growth at the edges.  Leaving ${\rm FoV\!}_{\rm NE}(z)$ as a generic function, and applying the same inverse square root scaling of this to derive the observed noise in a surveyed area, as above, gives:

\small\begin{eqnarray}
\bar{\sigma}_{S_{\nu}}(z) &=& f(z){\rm FoV\!}_{\rm NE}^{-\frac{1}{2}}(z) \bar{\sigma}_{S_{\nu}}(0) \,, \\
\bar{\sigma}_{S} (z) &=& f(z){\rm FoV\!}_{\rm NE}^{-\frac{1}{2}}(z) \bar{\sigma}_{S}(0) \,, \\
\bar{\sigma}_{S_{\nu}}^{\Delta V_{\rm rest}^{\rm const}} (z) &=& f(z){\rm FoV\!}_{\rm NE}^{-\frac{1}{2}}(z)  (1+z)^{{\frac{1}{2}}} \bar{\sigma}_{S}^{\Delta V_{\rm rest}}(0) \,, \\
\bar{\sigma}_{S}^{\Delta V_{\rm rest}^{\rm const}} (z) &=& f(z){\rm FoV\!}_{\rm NE}^{-\frac{1}{2}}(z) (1+z)^{{-\frac{1}{2}}} \bar{\sigma}_{S}^{\Delta V_{\rm rest}}(0) \,.
\end{eqnarray}\normalsize

\section{Brightness Temperature Sensitivity}

The relativistically derived relation between brightness temperature and flux density (Equation~\ref{eqn:TB}), gives a corresponding sensitivity relation of:

\begin{eqnarray}
\sigma_{T_{\rm B}} &=& (1+z)^{3}\frac{c^{2}}{2 k \nu_{\HIsub}^{2}\Omega_{\rm bm}} \sigma_{S_{\nu}} \,,
\end{eqnarray}

\begin{dmath}[breakdepth={1},style={\small}]
\Rightarrow \left(\frac{\sigma_{T_{\rm B}}}{\rm K}\right)\hiderel{=}6.86 \times 10^{5} \, (1+z)^{3} \left(\frac{\Omega_{\rm bm}}{\rm arcsec^{2}}\right)^{-1} \left(\frac{\sigma_{S_{\nu}}}{\rm Jy}\right) \,,
\end{dmath}

\noindent which combining with the previous expression for flux density sensitivity (Equation~\ref{eqn:sigsnu}) yields:

\begin{dmath}[breakdepth={3},style={\small}]
\sigma_{T_{\rm B}} = (1+z)^{3} \frac{c^{2}T_{\rm sys}}{\nu_{\HIsub}^{2}A_{e}\Omega_{\rm bm}} \frac{1}{\eta_{\rm s}\sqrt{2N(N-1)\Delta t \Delta \nu}} \,, \\
\end{dmath}

\begin{dmath}[breakdepth={0},style={\small}]
\Rightarrow \left(\frac{\sigma_{T_{\rm B}}}{\rm K}\right)\hiderel{=}\frac{1.34 \times 10^{9} \, (1+z)^{3}}{\eta_{\rm s}\sqrt{N(N-1)}} \times \left(\frac{\Omega_{\rm bm}}{\rm arcsec^{2}}\right)^{-1}  \left(\frac{A_{\rm e}/T_{\rm sys}}{\rm m^{2} K^{-1}}\right)^{-1} \left(\frac{\Delta t}{\rm s}\right)^{-\frac{1}{2}} \left(\frac{\Delta \nu}{\rm Hz}\right)^{-\frac{1}{2}} \,.
\end{dmath}
       
\noindent Using Equation~\ref{eqn:noiseSnu} this can also be expressed as a function of the flux density sensitivity per channel $\sigma_{\nu_{\rm chan}}$ (width $\Delta \nu_{\rm chan}$):

\begin{eqnarray}
\sigma_{T_{\rm B}} &=& (1+z)^{3} \frac{c^{2}}{2 k \nu_{\HIsub}^{2}\Omega_{\rm bm}} \sqrt{\frac{\Delta \nu_{\rm chan}}{\Delta \nu}}\sigma_{S_{\nu_{\rm chan}}}  \,,
\end{eqnarray}     

\begin{dmath}[breakdepth={1},style={\small}]
\Rightarrow \left(\frac{\sigma_{T_{\rm B}}}{\rm K}\right)\hiderel{=}6.86 \times 10^{5} \, (1+z)^{3} \times \sqrt{\frac{\Delta \nu_{\rm chan}}{\Delta \nu}} \left(\frac{\Omega_{\rm bm}}{\rm arcsec^{2}}\right)^{-1} \left(\frac{\sigma_{S_{\nu_{\rm chan}}}}{\rm Jy}\right) \,.
\end{dmath}
  
\noindent Or alternatively expressed per desired rest frame velocity interval $\Delta V_{\rm rest}$ rather than per observed frame frequency width using Equation~\ref{eqn:noiseSnuVrest}: 

\begin{eqnarray}
\sigma_{T_{\rm B}} &=& (1+z)^{7/2} \frac{c^{5/2}}{2 k \nu_{\HIsub}^{5/2}\Omega_{\rm bm}} \sqrt{\frac{\Delta \nu_{\rm chan}}{\Delta V_{\rm rest}}} \sigma_{S_{\nu_{\rm chan}}} \,,
\end{eqnarray}       
  
\begin{dmath}[breakdepth={1},style={\small}]
\Rightarrow \left(\frac{\sigma_{T_{\rm B}}}{\rm K}\right) \hiderel{=} 9971 \, (1+z)^{\frac{7}{2}} \times \left(\frac{\Omega_{\rm bm}}{\rm arcsec^{2}}\right)^{-1} \left(\frac{\Delta \nu_{\rm chan}}{\rm Hz}\right)^{\frac{1}{2}} \left(\frac{\Delta V_{\rm rest}}{\rm km\,s^{-1}}\right)^{-\frac{1}{2}} \left(\frac{\sigma_{S_{\nu_{\rm chan}}}}{\rm Jy}\right) \,.
\end{dmath}

\noindent In the instance of a comparison being made for observations with a given telescope at different redshifts, but where the configuration has been changed between observations to yield the same synthesised beam solid angle, this will give a redshift scaling relation of:

\begin{eqnarray}
\sigma_{T_{\rm B}}(z) &=& f(z)(1+z)^{3}  \sigma_{T_{\rm B}}(0) \,, \\
\sigma_{T_{\rm B}}^{\Delta V_{\rm rest}^{\rm const}}(z) &=& f(z)(1+z)^{7/2}  \sigma_{T_{\rm B}}^{\Delta V_{\rm rest}^{\rm const}}(0) \,,
\end{eqnarray}

\noindent where $f(z)$ captures redshift variations in $\eta_{\rm s}A_{\rm e}/T_{\rm sys}$ as before (Equation~\ref{eqn:ATscaling}).  If instead we now consider how the brightness temperature sensitivity will change for a fixed configuration observation across a range of frequencies, the synthesised beam solid angle will increase as $\Omega_{\rm bm}\propto (1+z)^{2}$ (ignoring frequency dependent illumination effects) giving:

\begin{eqnarray}
\sigma_{T_{\rm B}}(z) &=& f(z)(1+z)  \sigma_{T_{\rm B}}(0) \,, \\
\sigma_{T_{\rm B}}^{\Delta V_{\rm rest}^{\rm const}}(z) &=& f(z)(1+z)^{3/2}  \sigma_{T_{\rm B}}^{\Delta V_{\rm rest}^{\rm const}}(0) \,.
\end{eqnarray}
  
\noindent Taking these last equations as a starting point, we can also derive the redshift scaling equations for a widefield mosaic and for a phased array feed as per the previous section allowing for overlapping primary (formed-) beam effects:

\subsubsection*{Widefield Single Beam Mosaic:}
\begin{eqnarray}
\bar{\sigma}_{T_{\rm B}}(z) &=& f(z) \bar{\sigma}_{T_{\rm B}}(0) \,, \\
\bar{\sigma}_{T_{\rm B}}^{\Delta V_{\rm rest}^{\rm const}}(z) &=& f(z)(1+z)^{1/2}  \bar{\sigma}_{T_{\rm B}}^{\Delta V_{\rm rest}^{\rm const}}(0) \,.
\end{eqnarray}

\subsubsection*{Phased Array Feed Observations:}
\begin{eqnarray}
\bar{\sigma}_{T_{\rm B}}(z) &=& f(z){\rm FoV\!}_{\rm NE}^{-\frac{1}{2}}(z) \\ \nonumber && \times (1+z) \bar{\sigma}_{T_{\rm B}}(0) \,, \\
\bar{\sigma}_{T_{\rm B}}^{\Delta V_{\rm rest}^{\rm const}}(z) &=& f(z){\rm FoV\!}_{\rm NE}^{-\frac{1}{2}}(z)  \\ \nonumber && \times  (1+z)^{3/2} \bar{\sigma}_{T_{\rm B}}^{\Delta V_{\rm rest}^{\rm const}}(0) \,.
\end{eqnarray}     
                          
\section{Column Density Sensitivity}

\noindent Taking the previous expression for column density in terms of flux and beam size (Equation~\ref{eqn:nHI}) gives the corresponding sensitivity relation of: 

\begin{eqnarray}
\sigma_{N_{\HIsub}} &=& (1+z)^{4}  \frac{16\pi}{3 h\nu_{\HIsub} A_{\HIsub} \Omega_{\rm bm}}  \sigma_{S} \,,
\end{eqnarray}

\small\begin{eqnarray}
\Rightarrow  \left(\frac{\sigma_{N_{\HIsub}}}{\rm cm^{-2}}\right) &=& 2.64 \times 10^{20} \, (1+z)^{4}  \\ \nonumber && \times \left(\frac{\sigma_S}{\rm JyHz} \right) \left( \frac{\Omega_{\rm bm}}{\rm arcsec^{2}} \right) \,.
\end{eqnarray}\normalsize

\noindent and substituting in Equation~\ref{eqn:sigs} for a theoretical sensitivity estimate:

\small\begin{eqnarray}
\sigma_{N_{\HIsub}} &=& (1+z)^{4}  \frac{16\pi k T_{\rm sys}}{3 h\nu_{\HIsub} A_{\HIsub} \Omega_{\rm bm} \eta_{\rm s}A_{\rm e}}   \\ \nonumber && \times  \sqrt{\frac{2\Delta \nu}{N(N-1)\Delta t}} \,,
\end{eqnarray}\normalsize

\begin{dmath}[breakdepth={0},style={\small}]
\Rightarrow \left(\frac{\sigma_{N_{\HIsub}}}{\rm cm^{-2}}\right)\hiderel{=}\frac{5.15 \times 10^{23} \, (1+z)^{4}}{\eta_{\rm s}\sqrt{N(N-1)}} \times \left(\frac{\Omega_{\rm bm}}{\rm arcsec^{2}}\right)^{-1}  \left(\frac{A_{\rm e}/T_{\rm sys}}{\rm m^{2} K^{-1}}\right)^{-1} \left(\frac{\Delta t}{\rm s}\right)^{-\frac{1}{2}} \left(\frac{\Delta \nu}{\rm Hz}\right)^{\frac{1}{2}} \,.
\end{dmath}

\noindent Or from Equation~\ref{eqn:noiseS}, scaling from known channel noise:

\begin{eqnarray}
\sigma_{N_{\HIsub}} &=& (1+z)^{4} \frac{16\pi \sqrt{\Delta \nu_{\rm chan} \Delta \nu}}{3 h\nu_{\HIsub} A_{\HIsub} \Omega_{\rm bm}} \sigma_{S_{\nu_{\rm chan}}} \,,
\end{eqnarray}

\begin{dmath}[breakdepth={0},style={\small}]
\Rightarrow \left(\frac{\sigma_{N_{\HIsub}}}{\rm cm^{-2}}\right)\hiderel{=} 2.64 \hiderel{\times} 10^{20} \, (1+z)^{4} \times\left(\frac{\Omega_{\rm bm}}{\rm arcsec^{2}}\right)^{-1}  \left(\frac{\Delta \nu}{\rm Hz}\right)^{\frac{1}{2}} \left(\frac{\Delta\nu_{\rm chan}}{\rm Hz}\right)^{\frac{1}{2}} \left(\frac{\sigma_{S_{\nu_{\rm chan}}}}{\rm Jy}\right) \,.
\end{dmath}

\noindent And finally as a function of rest frame velocity interval using Equation~\ref{eqn:noiseSVrest}:

\begin{eqnarray}
\sigma_{N_{\HIsub}} &=& (1+z)^{7/2}  \frac{16\pi \sqrt{\Delta V_{\rm rest}\Delta\nu_{\rm chan}}}{3 h\nu_{\HIsub}^{1/2} A_{\HIsub} \Omega_{\rm bm}c^{1/2}}  \sigma_{S_{\nu_{\rm chan}}} \,, \label{eqn:sigmanHIVrest} 
\end{eqnarray}

\begin{dmath}[breakdepth={0},style={\small}]
\Rightarrow \left(\frac{\sigma_{N_{\HIsub}}}{\rm cm^{-2}}\right)\hiderel{=} 1.82 \hiderel{\times} 10^{22} \, (1+z)^{\frac{7}{2}} \times\left(\frac{\Omega_{\rm bm}}{\rm arcsec^{2}}\right)^{-1}  \left(\frac{\Delta V_{\rm rest}}{\rm km\,s^{-1}}\right)^{\frac{1}{2}} \left(\frac{\Delta\nu_{\rm chan}}{\rm Hz}\right)^{\frac{1}{2}} \left(\frac{\sigma_{S_{\nu_{\rm chan}}}}{\rm Jy}\right) \,.
\end{dmath}

\noindent Looking at the redshift scaling relations as before, we have for a changing configuration with fixed synthesised beam solid angle (with $f(z)$ defined as before in Equation~\ref{eqn:ATscaling}):

\begin{eqnarray}
\sigma_{N_{\HIsub}}(z) & = & f(z) (1+z)^{4} \,\, \sigma_{N_{\HIsub}}(0) \,, \\ 
\sigma_{N_{\HIsub}}^{\Delta V_{\rm rest}^{\rm const}}(z) & = & f(z) (1+z)^{7/2} \,\, \sigma_{N_{\HIsub}}^{\Delta V_{\rm rest}^{\rm const}}(0) \,.
\end{eqnarray}

\noindent For a fixed configuration with synthesised beam solid angle increasing as $(1+z)^{2}$:

\begin{eqnarray}          
\sigma_{N_{\HIsub}}(z) & = & f(z) (1+z)^{2} \,\, \sigma_{{}N_{\HIsub}}(0) \,, \\  
\sigma_{N_{\HIsub}}^{\Delta V_{\rm rest}^{\rm const}} (z) & = & f(z) (1+z)^{3/2} \,\, \sigma_{N_{\HIsub}}^{\Delta V_{\rm rest}^{\rm const}} (0) \,.
\end{eqnarray}

\noindent And finally for the widefield mosaic and phased array feed cases:
 
\subsubsection*{Widefield Single-Beam Mosaic:}

\begin{eqnarray}
\bar{\sigma}_{N_{\HIsub}}(z) & = & f(z) (1+z) \,\, \bar{\sigma}_{N_{\HIsub}}(0)  \,, \\
\bar{\sigma}_{N_{\HIsub}}^{\Delta V_{\rm rest}^{\rm const}} (z) & = & f(z) (1+z)^{1/2} \,\, \bar{\sigma}_{N_{\HIsub}}^{\Delta V_{\rm rest}^{\rm const}} (0) \,.
\end{eqnarray}

\subsubsection*{Phased Array Feeds Observations:}

\begin{eqnarray}
\bar{\sigma}_{N_{\HIsub}}(z) & = & f(z) {\rm FoV\!}_{\rm NE}^{-\frac{1}{2}}(z)  \\ \nonumber && \times (1+z)^{2} \,\, \bar{\sigma}_{N_{\HIsub}}(0) \,,  \\
\bar{\sigma}_{N_{\HIsub}}^{\Delta V_{\rm rest}^{\rm const}} (z) & = & f(z) {\rm FoV\!}_{\rm NE}^{-\frac{1}{2}}(z)  \\ \nonumber && \times (1+z)^{3/2} \,\, \bar{\sigma}_{N_{\HIsub}}^{\Delta V_{\rm rest}^{\rm const}}(0) \,. \\
\end{eqnarray}

\section{Signal-to-Noise}

The significance of a measurement can be determined by simply taking the ratio of the observed quantity to the measured noise in that quantity over the same region.  As such, the equations of the previous sections trivially lend themselves to predicting the likely outcomes of observations.  A few of the most important are detailed below.\\

\noindent Peak signal-to-noise:

\begin{eqnarray} 
S/N_{\rm peak} &=& \frac{S_{\rm \nu,peak}}{\sigma_{S_{\nu_{\rm chan}}}} \,.
\end{eqnarray}

\noindent Integrated signal-to-noise for a point source:

\begin{eqnarray} 
S/N_{\rm int} &=& \frac{S}{\sigma_{S_{\nu_{\rm chan}}}\sqrt{\Delta \nu_{\rm chan} \Delta \nu}} \,, \\
                  &=& \frac{S}{\sigma_{\rm chan}\Delta \nu_{\rm chan} \sqrt{N_{\rm chan}}} \,,
\end{eqnarray}

\noindent where the width of the profile is specified as either a frequency units ($\Delta \nu$) or the number of channels ($N_{\rm chan}$).  It is worth noting in the above equations that the product $\sigma_{\rm chan}\sqrt{\Delta_{\nu_{\rm chan}}}$ is independent of the particular choice of channel width used, and as such $S/N_{\rm int}$ is independent of channel choice. To predict the S/N for a point source of known mass and rest frame velocity width, the above can also be alternatively expressed:

\begin{eqnarray} 
S/N_{\rm int} &=& \frac{S}{\sigma_{S_{\nu_{\rm chan}}}}\sqrt{\frac{c(1+z)}{\nu_{\HIsub} \Delta \nu_{\rm chan} \Delta V_{\rm rest}}} \,, \\
\end{eqnarray}

\noindent which substituting for $S$ using Equation.~\ref{eqn:HImass} gives:

\begin{dmath}[breakdepth={3},style={\small}]
S/N_{\rm int} = 2.92\times10^{-4} (1+z)^{\frac{1}{2}}   \left( \frac{D_L}{\rm Mpc} \right)^{-2}  \times\left( \frac{M_{\HIsub}}{M_{\odot}} \right)  \left( \frac{\Delta V_{\rm rest}}{\rm km\,s^{-1}} \right)^{-\frac{1}{2}} \left( \frac{\Delta \nu_{\rm chan}}{\rm Hz} \right)^{-\frac{1}{2}}     \left( \frac{\sigma_{S_{\nu_{\rm chan}}}}{\rm Jy}  \right)^{-1} \,.
\end{dmath}

\noindent If the source is extended, a further correction of $S/N_{\rm int} \propto 1/\sqrt{(1 + A_{\rm galaxy}/A_{\rm beam}}$ is required to allow for the fact that source flux is distributed over multiple beams, increasing the observed noise \citep{duffy2012}.  Note that this correction factor only disappears if the source is a true point source, rather than just being smaller than the (synthesised) beam.  The estimation of signal-to-noise using this correction factor also assumes the optimal extraction of the source signal in both spatial and frequency coordinates, which for low signal-to-noise sources may not be possible.

For an extended source, a generally more appropriate calculation is to determine the expected signal-to-noise from the desired column density sensitivity.  For instance, considering the case of single pointing with beam $\Omega_{\rm bm}$ at redshift $z$, the signal-to-noise in the centre of the primary beam will be given by (using Equation~\ref{eqn:sigmanHIVrest}):

\begin{dmath}[breakdepth={1},style={\small}]
S/N_{N_{\HIsub}} = \frac{N_{\HIsub_{\rm desired}}}{\sigma_{N_{\HIsub}}^{\Delta V_{\rm rest}}} \,, \\
\simeq 5.50 \times 10^{-23}  \, (1+z)^{\frac{2}{7}}  \left(\frac{N_{\HIsub_{\rm desired}}}{\rm cm^{-2}}\right) \times \left(\frac{\Omega_{\rm bm}}{\rm arcsec^{2}}\right)  \left(\frac{\Delta V_{\rm rest}}{\rm km\, s^{-1}}\right)^{-\frac{1}{2}}   \left(\frac{\Delta\nu_{\rm chan}}{\rm Hz}\right)^{-\frac{1}{2}}  \left( \frac{\sigma_{S_{\nu_{\rm chan}}}}{\rm Jy}  \right)^{-1}   \,.
\end{dmath}

\section{HI Fidelity Calculator}
\label{sec:hifi}

As a complementary online tool, the HI Fidelity (HiFi) calculator makes available many of the central formulae identified in this paper for ready application to the analysis of, or planning for, HI observations (\url{http://hifi.icrar.org}). These include the conversion of observed frame quantities to rest frame equivalents and vice versa, including: frequency width and velocity width, flux and mass, flux density and brightness temperature, and flux and column density.  Also included are calculators for the estimation of observed noise and signal-to-noise, combining observed frame measurement characteristics with rest frame source properties.  

The calculator makes use of the Celestial R package\footnote{\url{adsabs.harvard.edu/abs/2016ascl.soft02011R, cran.r-project.org/web/packages/celestial/index.html}} for the calculation of cosmological distances.  In addition to the non-evolving vacuum energy density calculation outlined in Section~\ref{sec:distances}, this package, and the HiFi calculator, also allow for an evolving vacuum energy model through the use of the $w$ and $w'$ parameters, replacing $\Omega_{\Lambda}$ in Equation~\ref{eqn:Ez} with: 

\begin{equation}
\Omega_{\Lambda} \rightarrow \Omega_{\Lambda}\left( \frac{1}{1+z} \right) ^{-(3+3w_{0}+6w')} e^{ -6w' \left( 1 - \frac{1}{1+z} \right) } \,,
\end{equation}

\noindent where $w_{0} =-1$ and $w'=0$ return the default non-evolving vacuum energy density.

\section*{Acknowledgments} 

Parts of this research were conducted by the Australian Research Council Centre of Excellence for All-sky Astrophysics (CAASTRO), through project number CE110001020.


\nocite*{}
\bibliographystyle{pasa-mnras}
\bibliography{hifi.bib}

\section*{Appendix A: litte h conversions}

\noindent If you need to determine the value of some quantity for a different Hubble constant, and the $h_{\rm C}$ dependencies are given, this can be done by simply calculating the value of $h_{\rm C}$ for your new Hubble constant and then evaluating the expression given for the quantity. For example, to calculate the value of an HI mass given as $M_{\HIsub} = 10^{9.7} h_{73}^{-2} M_{\odot}$ for a Hubble constant of $H_{\rm 0} = 100$:

\begin{eqnarray}
h_{73} &=&  \frac{H_{\rm 0}}{73\,\, \mathrm{km\, s^{-1} Mpc^{-1}}} \,, \\
            &=&  \frac{100\,\, \mathrm{km\, s^{-1} Mpc^{-1}}}{73\,\, \mathrm{km\, s^{-1} Mpc^{-1}}} \,, \\
            &=&  1.37 \,.
\end{eqnarray}

\noindent then, 

\begin{eqnarray}
M_{\HIsub} &=& 10^{9.7} h_{73}^{-2} M_{\odot} \,, \\
	&=& 10^{9.7} 1.37^{-2} h_{100}^{-2} M_{\odot} \,, \\
	&=& 10^{9.4} h_{100}^{-2} M_{\odot} \,.
\end{eqnarray}

\noindent  Alternatively running this in reverse, and this time using the abbreviation $h = h_{100}$:

\begin{eqnarray}
M_{\HIsub} &=& 10^{9.4} h^{-2} M_{\odot} \,, \\
            &=& 10^{9.4} {0.73}^{-2} h_{73}^{-2} M_{\odot} \,, \\
	   &=& 10^{9.7} h_{73}^{-2} M_{\odot} \,.
\end{eqnarray}

\noindent A final point to note is that the little h dependencies for a quantity can be different, depending on how it was determined.  A classic example being the differences that often arise in little h exponents between observed and simulated quantities.  The important thing to do if you are wanting to compare two such quantities is to just make sure both values are valid for the same Hubble constant, rather than worrying about the exponents of $h_{\rm C}$, which are inherently different.\\

\section*{Appendix B: Symbols, Units, Constants \& Glossary}

\noindent Units:

\begin{eqnarray}
1 \,\rm{Jy} &=& 1^{-26} \,\rm{W} \,\rm{m}^{-2} \,\rm{Hz}^{-1}\\
M_{\odot} &=& 1.98855 \times 10^{30} {\, \rm kg}
\end{eqnarray}

\noindent HI constants used in this work:
\begin{eqnarray}
\nu_{\HIsub} &=& 1.420405751786\times 10^{9} \, \rm{\,Hz}\\ 
A_{\HIsub} &=& 2.86888\times10^{-15} {\,\rm s^{-1}}\\ 
m_{\rm H} &=& 1.673533 \times 10^{-27} {\,\rm kg}
\end{eqnarray}

\noindent Glossary of quantities used in this paper and their units:

\begin{eqnarray}
\nu &=& \mbox{frequency} \nonumber \\
	&&\mbox{(Hz)}\nonumber \\
V &=& \mbox{velocity} \nonumber \\
	&&\mbox{(km$\,$s$^{-1}$)}\nonumber \\
S_{\nu} &=& S_{\nu_{\rm obs}} = \mbox{received flux density} \nonumber \\ 
	  && \mbox{(10$^{-26}$ W m$^{-2}$ Hz$^{-1}$ = Jy)} \nonumber \\
S &=& S_{\rm obs} \nonumber = \mbox{received flux} \nonumber \\
      && \mbox{(10$^{-26}$ W m$^{-2}$ = Jy Hz)} \nonumber \\
S^{V} &=& S^{V}_{\rm obs} = \mbox{received velocity integrated flux} \nonumber \\
	  &&  \mbox{(10$^{-26}$ W m$^{-2}$ Hz$^{-1}$ km s$^{-1}$ = Jy km s$^{-1}$)} \nonumber \\
L_{\nu} &=& L_{\nu_{\rm rest}}  = \mbox{emitted luminosity density} \nonumber \\
	&&\mbox{(W Hz$^{-1}$)}\nonumber \\
L &=& L_{\rm rest} = \mbox{emitted luminosity} \nonumber \\
	&& \mbox{(W)} \nonumber \\
I_{\nu} &=& I_{\nu_{\rm obs}} = \mbox{received specific intensity density} \nonumber\\
            &&  \mbox{(W m$^{-2}$ sr$^{-1}$ Hz$^{-1}$)} \nonumber \\
I &=& I_{\rm obs} = \mbox{received specific intensity} \nonumber \\
	&& \mbox{(W m$^{-2}$ sr$^{-1}$)} \nonumber \\
B_{\nu} &=& B_{\nu_{\rm rest}} = \mbox{emitted surface brightness density} \nonumber \\
	&& \mbox{ (W m$^{-2}$ sr$^{-1}$ Hz$^{-1}$)} \nonumber \\
B &=& B_{\rm rest} = \mbox{emitted surface brightness} \nonumber \\
	&& \mbox{(W m$^{-2}$ sr$^{-1}$)} \nonumber \\
\mathscr{N}_{\HIsub} &=& \mbox{number of HI atoms} \nonumber \\
N_{\HIsub} &=& \mbox{column density of HI atoms (cm$^{-2}$)}  \nonumber \\
n_{\HIsub} &=& \mbox{volume density of HI atoms (cm$^{-3}$)}  \nonumber 
\end{eqnarray}

\end{document}